\documentclass[twocolumn,secnumarabic,amssymb, nobibnotes, aps, prl, reprint]{revtex4-1}

\usepackage{amsmath}
\usepackage{graphicx}
\usepackage{bm}
\usepackage[usenames,dvipsnames]{color} 

\usepackage[version=3]{mhchem} 

\begin{document}



\title{Non-decaying hydrodynamic interactions along narrow channels}

\author{Karolis Misiunas}
\email{karolis@misiunas.com}
\author{Stefano Pagliara}
\thanks{Present address: Department of Biosciences, College of Life and Environmental Sciences, University of Exeter, Exeter, United Kingdom}
\author{Eric Lauga}
\author{John R. Lister}
\author{Ulrich F. Keyser}
\email[correspondence: ]{ufk20@cam.ac.uk}
\affiliation{Cavendish Laboratory, University of Cambridge, UK}
\affiliation{Department of Applied Mathematics and Theoretical Physics , University of Cambridge, UK}
\date{\today}%

\begin{abstract}
    Particle-particle interactions are of paramount importance in every multi-body system as they determine the collective behaviour and coupling strength. Many well-known interactions like electro-static, van der Waals or  screened Coulomb, decay exponentially or with negative powers of the particle spacing $r$. Similarly, hydrodynamic interactions between particles undergoing Brownian motion decay as $1/r$ in bulk, and are assumed to decay in small channels. Such interactions are ubiquitous in biological and technological systems. Here we confine two particles undergoing Brownian motion in narrow, microfluidic channels and study their coupling through hydrodynamic interactions. Our experiments show that the hydrodynamic particle-particle interactions are distance-independent in these channels. This finding is of fundamental importance for the interpretation of experiments where dense mixtures of particles or molecules diffuse through finite length, water-filled channels or pore networks.
\end{abstract}

\maketitle


Molecular diffusion inside channels and pores is relevant for a wide range of phenomena in natural systems~\cite{alberts2008molecular} as well as technological applications~\cite{Karger2014,Motz2014}. For example, biological channels that transport antibiotics are under intense investigation due to their importance for drug transport~\cite{Pages2008}. Diffusing molecules inside protein channels are closely confined, leading to single-file diffusion with hydrodynamic interactions playing a role even at these nanometre length-scales~\cite{Deen1987,Gravelle2013}. Transport through protein channels can be mimicked by colloidal particles and microfluidics chips~\cite{Pagliara2014} that confine particles to 1D diffusion~\cite{Wei2000}. Such particles undergo random walks in one dimension driven by Brownian motion, where the corresponding diffusion coefficient critically depends on the geometry of the confinement~\cite{Dettmer2014a,Motz2014}.

Loosely speaking, these Brownian particles receive momentum impulses from thermal fluctuations of the solvent molecules~\cite{Huang2010a}, and their resulting motion displaces the liquid around it~\cite{Deen1987,Happel1973}. This creates a flow field that mediates a long-range hydrodynamic interaction between the particles~\cite{Diamant2009}. The strength of this interaction is proportional to the flow velocity which decays with distance as $1/r$, where $r$ is the separation between two unconstrained particles in three dimensions (3D)~\cite{Crocker1997}. 
Introducing a geometrical confinement modifies the flow field that changes the decay rate. For example, the interaction between particles constrained by two parallel plates (2D) decays faster at a rate of $\sim 1/r^2$~\cite{Cui2004}.

In narrow channels the hydrodynamic interactions were previously measured to rapidly decay with particle separation~\cite{Cui2002,Valley2007a,Kosheleva2012}. Indeed, the steady flow induced by particle motion in a channel decays exponentially with $r/2R$, where $2R$ is the channel width~\cite{Happel1973,Deen1987,Liron1978,AlQuddus2008}. Consequently, the particle-particle interaction strength is expected to also decay exponentially with their separation~\cite{Liron1978,Cui2002}. 
However, recent theoretical investigation suggests that interactions could have a longer spacial extent and a slower decay rate than previously thought~\cite{Frydel2010}.
Furthermore, previous experiments did not capture far field hydrodynamics  because they used microfluidic chips with a groove geometry for mimicking channels~\cite{Wei2000,Cui2002,Valley2007a,Kosheleva2012}. Sedimentation kept the particles from escaping the microfluidic groove, but the liquid had no such constraint. The lack of controlled experiments in 1D confinement leaves many unanswered questions about the magnitude and spacial extent of hydrodynamic interactions inside narrow channels~\cite{Wei2000,Reguera2006}.

\begin{figure}[ht!]
    \centering
    \includegraphics[width=8cm]{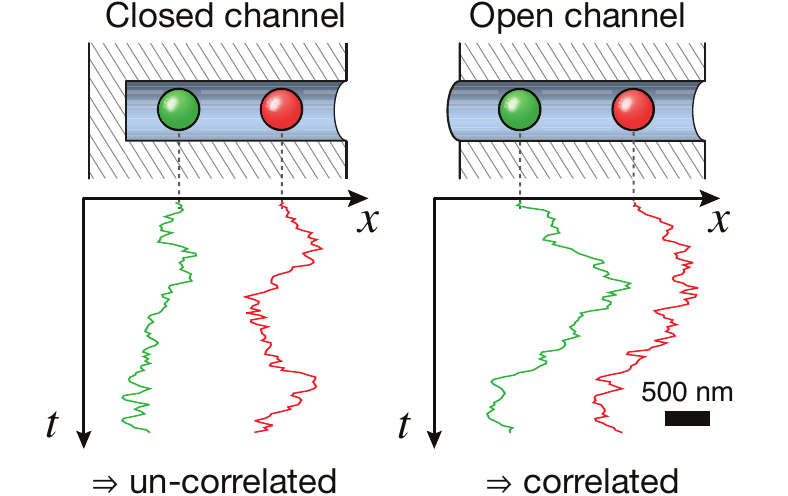}
    \caption{\label{illustation_closed_and_open_channel_correlation}
    Trajectories of two particles undergoing Brownian motion in closed and open channels. The trajectories are visibly un-correlated in the closed channel, suggesting the particles move independently. In the open channel the trajectories resemble each other, which leads to strong motion correlation.
    }
\end{figure}

In this paper, we present the first measurement of the interactions between two Brownian particles inside a finite narrow channel, that confines both the particles and the liquid. Figure~\ref{illustation_closed_and_open_channel_correlation} illustrates our experiments, where the `closed' channel, shown on the left, has only one end connected to a bulk reservoir, while the `open' channel, shown on the right, has both ends connected. Here we will demonstrate a fundamental difference between diffusion in open channels and in closed channels~\cite{Happel1973,Cui2002,Liron1978}.

\begin{figure}[ht!]
    \centering
    \includegraphics[width=8cm]{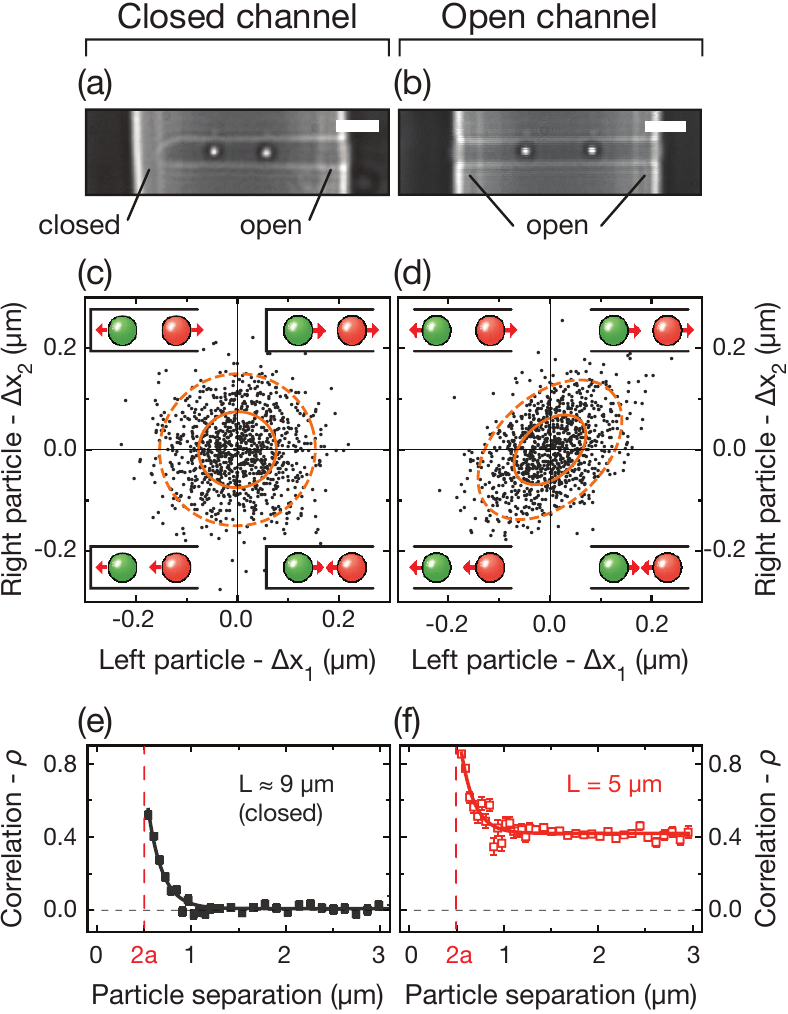}
    \caption{\label{jump_map_and_correlation_single_column}
    Comparison between a closed channel (left column) and an open channel (right column) experiments.
    The top row (a,b) shows bright field images of microfluidic channels containing two colloidal particles. Scale bars indicate $2~\mu$m.
    The middle row (c,d) shows displacements of the second particle ($\hat y$ axis) as a function of displacement of the first particle ($\hat x$ axis). Each distribution contains $1000$ displacement pairs with an initial particle separation of approximately $2.5\;\mu$m.
    The overlaid lines indicate contours for $\sigma$ and $2\sigma$ from a 2D normal-distribution fit. Insets illustrate the direction of motion for each quadrant.
    The bottom row (e,f) shows the correlation coefficients versus the separation between the two particles. The solid lines indicate the fits to the phenomenological model.
    The two-particle motion is strongly correlated in the open channel, suggesting a presence of long-ranged non-decaying interaction.
    }
\end{figure}

The experiments are realised using microfluidic lab-on-a-chip devices because they allow for direct imaging of particle motion. Inside the chip two large reservoirs are separated by a membrane containing closed (figure~\ref{jump_map_and_correlation_single_column}a) and open channels (figure~\ref{jump_map_and_correlation_single_column}b). All channels are $5-17\;\mu$m long and have semi-elliptical cross-sections with a height and width of approximately $800\;$nm that closely confine spherical particles of diameter ${2a = 505 \pm 8\;}$nm.
The resulting particle to channel size ratio is ${a/R \approx 0.6}$, ensuring  particles always move in single-file. Two additional large connections are positioned ${{\sim}200\;\mu}$m away from the narrow channels allowing pressure equalisation between the two reservoirs. For the fabrication of the chips, we use focused ion beam, photo-lithography and replica moulding of polydimethylsiloxane (PDMS)~\cite{Pagliara2011}. Crucially, the PDMS chips are then oxygen plasma bonded onto a glass slide which provides a bottom wall confinement for the channels. Subsequently, we fill the chip with the polystyrene colloidal particles (Polysciences Inc.) dispersed in a $5$~mM~\ce{KCl} salt solution that limits electrostatic interactions to a few nanometers.

An assembled chip is mounted onto an inverted, custom-built optical microscope with a high numerical aperture oil-immersion objective ($\times$100; NA~1.4; UPLSAPO). Using holographic optical tweezers~\cite{Grier2003,Padgett2011,Pagliara2013} we position two particles inside the channel, and then turn off the laser trapping which releases the particles and allows them to diffuse freely. Their motion is recorded using a CMOS camera (DMK-31BF03, Imaging Source) at a rate of $30$~frames per second, until one particle escapes the channel. Afterwards, the particle trajectories are extracted from the images using standard image analysis techniques~\cite{Dettmer2014}. See supplementary videos S1 and S2.


For data analysis, we divide the trajectories into displacement steps between consecutive frames: $\Delta x_1$ and $\Delta x_2$ denote the displacements of the first particle (one on the left) and the second particle (one on the right), respectively. The two-particle interaction strength is expected to be a function of their separation, and therefore, we group the pairs of displacements according to distance between centres of the particles.

For the closed channel, figure~\ref{jump_map_and_correlation_single_column}c shows the displacements of the second particle (${\Delta}x_2$) as a function of displacement of the first particle (${\Delta}x_1$). The distribution is circular with points distributed equally in each quadrant, suggesting the two particles move independently from each other. This is expected for a large particle separation of $10a$~\cite{Liron1978,Cui2002}.  
In contrast, the distribution for the open channel, shown in figure~\ref{jump_map_and_correlation_single_column}d, is elliptical with the major axis along $y=x$ and ellipticity of $0.75$. This implies that the two particles move in the same direction more frequently than in opposite directions, suggesting the presence of long-ranged interaction between the particles.

We quantify the interaction strength using the Pearson product-moment correlation coefficient defined as~\cite{riley2006mathematical}: ${\rho\,=\,\mathrm{cov}(\Delta{}x_1,\,\Delta{}x_2)\,/\,\sigma(\Delta{}x_1)\,\sigma(\Delta{}x_2)}$, where $\mathrm{cov}$ is the covariance, and $\sigma$ is the standard deviation. The value $\rho=0$ indicates independent particle motion and $\rho=1$ corresponds to fully correlated motion.

Figures~\ref{jump_map_and_correlation_single_column}e and \ref{jump_map_and_correlation_single_column}f show the correlation coefficients as a function of particle separation. Evidently, the correlation is stronger and has a longer range in the open channel. A detailed examination of the closed channel results shows a high correlation when particles are close to one another that exponentially decays to zero in a separation of ${\sim}4a$, as expected.
Therefore, we fit a phenomenological model, $y=A\,\exp(-x/B)+C$, that captures the decay rate and has an additional offset parameter. The fit yields $B=0.15 \pm .02\;\mu$m, $C=0.005 \pm .004$. We have introduced the offset to characterise the novel behaviour observed in the open channel.
As evident from the data, the correlation coefficient exhibits the same initial exponential decay, but in stark contrast, it asymptotes to a constant offset. This finite correlation coefficient is captured by the fit to the phenomenological model, yielding $B=0.14 \pm .02\;\mu$m, $C=0.419 \pm .005$. This constant, non-decaying component is not expected. Furthermore, the correlation persists even at the largest measured distances, suggesting that the two particles interact when both are inside the open channel. To the best of our knowledge, this is the first observation of such distance-independent interactions between Brownian particles.


\begin{figure}[!ht]
    \centering
    \includegraphics[width=8cm]{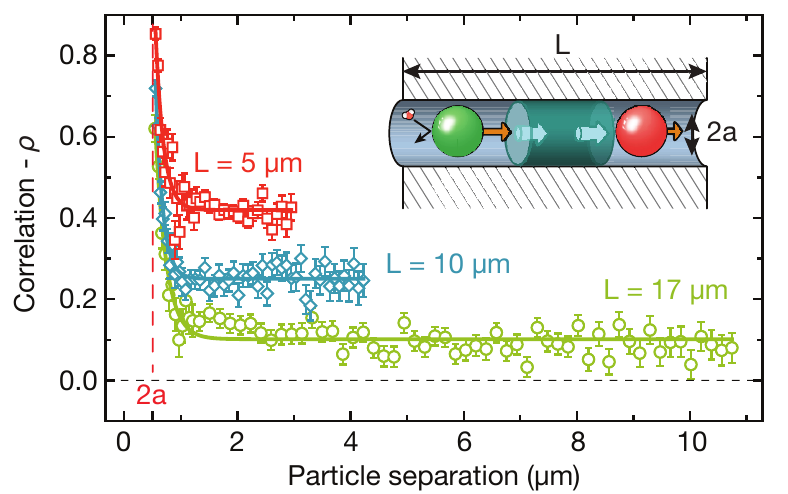}
    \caption{\label{correlation_VS_different_channels}
    Correlation coefficients for two interacting particles for different channel lengths. The three curves correspond to open channels with different lengths, from top to bottom: $5~\mu$m, $10~\mu$m, $17~\mu$m. The solid lines show the fit to a model $y=A\,\exp(-x/B) + C$, where the offset coefficients were (from top to bottom): $C=0.420 \pm .006$; $0.250 \pm .005$; $0.101 \pm .004$.
    Evidently, the long-ranged two-particle correlation coefficient decreases with the channel length. Inset illustrates the proposed model for long-ranged hydrodynamic interaction. }
\end{figure}

We further investigate the effect of the channel length on the particle-particle interaction strength. Figure~\ref{correlation_VS_different_channels} shows the correlation coefficients for open channels of lengths $L=5,\;10$ and $17\,\mu$m. The data clearly indicates that the interaction strength decreases with $L$. In the longest channel, shown as circles, we observe particles interacting at separations of more than $40a$. This is the largest relative distance measured between interacting Brownian particles, even exceeding the ${\sim}20a$ separation measured in bulk~\cite{Crocker1997,Dufresne2000}. This is a surprising result because the geometric confinement typically reduces the maximum interaction distance~\cite{Dufresne2000,Happel1973,Cui2002}.


Based on our observations we propose a hypothesis for the distance-independent interaction mechanism, that is schematically illustrated in the inset of figure~\ref{correlation_VS_different_channels}.
Suppose the first particle, shown on the left, moves to the right due to a thermal momentum impulse.
At steady state its motion induces a flow that is constrained by the channel geometry to flow either around the particle or along the channel. Importantly, the latter flow has been previously neglected~\cite{Liron1978,Cui2002,Happel1973,Deen1987} because the studies considered infinitely long channels, in which the finite pressure exerted by the moving particle cannot displace an infinite liquid column. In contrast, we argue that for finite open channels flows extend throughout the whole channel. Consequently, the induced flow along the channel has a constant mean flow velocity that is proportional to the driving force, i.e.\ the first particle's velocity~\cite{Happel1973,Mazo2002}. This flow encounters the second particle, shown on the right, and exerts a viscous drag force on it. The magnitude of this force is a function of flow velocity~\cite{Smythe1961} and thus also a function of the first particle's velocity, but is independent of the particle positions.
The opposite case of the second particle moving can be inferred by symmetry, and the final interaction is a combination of the two cases. This gives rise to the particle-particle interaction that we observed with our correlation coefficient measurements and also explains why the value is distance-independent in open channels. In a limit, where the particle size matches the channel, this problem resembles two pistons in a pipe and the trajectories would be perfectly correlated~\cite{Diamant2009}. On the contrary, in the closed channels, the dead end blocks the flow along the channel, thus eliminating the non-decaying hydrodynamic interaction.


We now use this hypothesis to construct an analytical model that sheds light into the physics of the interaction. The model focuses on physical scalings and omits detailed numerical pre-factors from the presentation. Our goal is to estimate the typical mean flow velocity in the channel resulting from the motion of one particle.

Consider a spherical particle of diameter $2a$ located in the centre of an open cylindrical channel of radius $R$ and length $L$. We assume that the particle moves instantaneously to the right with velocity $U$. The fluid displaced by the particle must either be pushed along the channel to the right channel end, with more fluid drawn in at the left, or leak from right to left through the thin gap between the particle and the channel wall. The pressure increase across the sphere, $\Delta p$, is proportional to the flow rate in the channel, $Q$, according to Poiseuille law

\begin{equation}\label{1}
    \Delta p \sim \frac{\mu Q L}{R^4},
\end{equation}

\noindent
where $\mu$ is the dynamic viscosity. We neglect any hydrodynamic resistance due to the recirculation from the exit of the channel to the entrance, equivalent to imposing a periodic boundary condition. Mass conservation around the moving sphere (in the frame moving with the sphere) leads to

\begin{equation}\label{2}
    Q - U R^2 \sim R q,
\end{equation}

\noindent
where $q$ is the leakage flux through the thin gap between the sphere and the channel. It is approximately given by lubrication theory~\cite{batchelor2000introduction}

\begin{equation}\label{3}
    q \sim - U h - \frac{\Delta p h^3}{\mu l_0},
\end{equation}

\noindent
where $h\equiv R-a$ is the minimum gap width, and ${l_0 \sim (ah)^{1/2}}$ is the characteristic lubrication length-scale. Combining these equations and taking the limit $h\,\ll\,R$, we obtain the flow rate

\begin{equation}
    Q \sim U R^2
        \left( \frac{R^3  a^{1/2} }{R^3  a^{1/2}+ L  h^{5/2}   }\right).
\end{equation}

Combining with the mean flow in the channel, ${\langle u \rangle \sim Q/R^2}$, we get the final expression

\begin{equation}
    \frac{\langle u\rangle }{U}\sim \frac{R^3  a }{R^3a  + L\;  h^{5/2}a^{1/2}}\cdot
\end{equation}

Our result shows that the mean flow in the channel is non-zero, but a function of the channel dimensions $R$ and $L$. Importantly, our model reproduces the observed decrease with the channel length - scaling with $1/L$. The long channel limit (${L \to \infty}$) gives no external flows ($Q=0$) equivalent to the closed channel case.


In order to quantitatively capture the experimental results, we employ numerical simulations. Using finite element analysis software (COMSOL Multiphysics v4.4)~\cite{Dettmer2014a} we solve the Stokes equations for two spherical particles inside a cylindrical channel with a diameter chosen such that the areas match the experiment parameters. Periodic boundary conditions are set on the ends while all other surfaces were set to no-slip boundary conditions. We apply an instantaneous velocity on one particle and calculate the resulting drag forces on both particles. A velocity independent measure of these forces is known as a friction matrix ($\boldsymbol{\zeta}$)~\cite[p. 226]{Mazo2002}. For the two particle system in 1D it is a $2\,{\times}\,2$ symmetric matrix, where the diagonal terms ($\zeta_{1,1};\; \zeta_{2,2}$) describe hydrodynamic drag experienced by the moving particles (first; second), and the off-diagonal elements ($\zeta_{1,2}=\zeta_{2,1}$) correspond to a force exerted by the moving particle on the other particle.
The friction matrix values are computed directly from the numerical simulation. In addition, the friction matrix can be obtained from the experiments using a diffusivity matrix, ${\boldsymbol{D} = k_B T \boldsymbol{\zeta}^{-1}}$~\cite{Mazo2002}, with components  ${D_{i,i} = \langle \Delta{}x_i^2 \rangle / 2 \Delta{}t }$ and ${D_{1,2} = D_{2,1} = \langle \Delta{}x_1 \, \Delta{}x_2 \rangle / 2 \Delta{}t }$~\cite{Xu2005}. This allows us to compare the hydrodynamic interactions predicted by our simulation with the measured data.

\begin{figure*}[!ht]
    \centering
    \includegraphics[width=13.6cm]{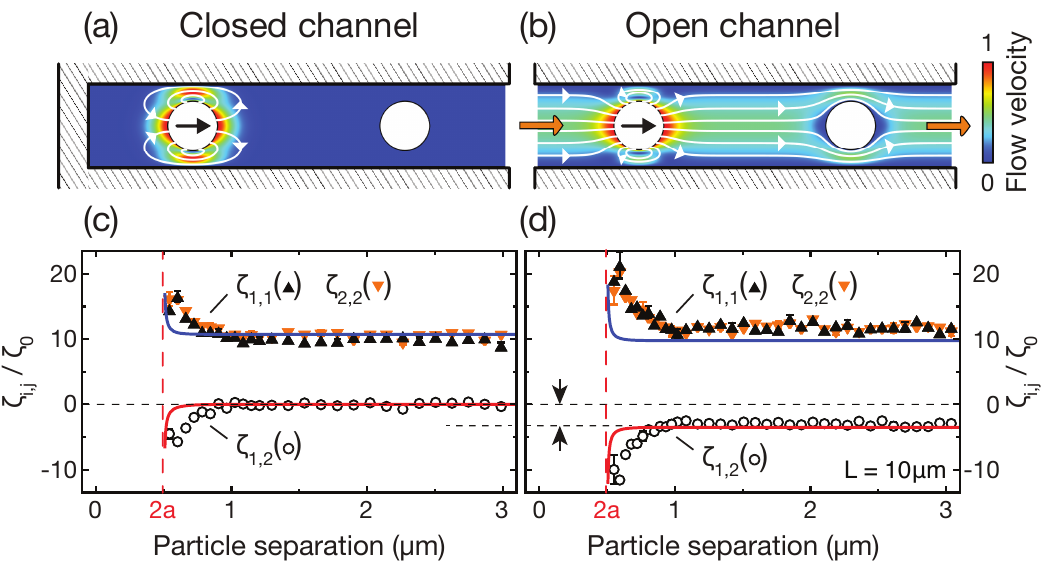}
    \caption{\label{friction_coefficients_for_experiments_and_simulations}
    A comparison between experimental data and simulation results.
    Top row illustrates flows inside the (a) closed and (b) open channel based on simulation results.
    The proportions were chosen for illustrative purposes.
    The bottom row compares the friction matrix values as function of particle separation for the (c) closed and (d) open channel. The points indicate experimentally measured values, while solid lines show the corresponding simulation results obtained without any fitting parameters. All values are normalised with the friction coefficient in the bulk: ${\zeta_0 = 6 \pi \mu a}$. Error bars are shown on every fifth data point.
    Arrows highlight the non-zero interaction term which agrees well with our simulation results, suggesting that the non-decaying interaction is caused by the induced flow inside the channel.
    }
\end{figure*}

Figures~\ref{friction_coefficients_for_experiments_and_simulations}a and~\ref{friction_coefficients_for_experiments_and_simulations}b illustrate the typical flow patterns computed with our simulation.
Notice that the flow in the closed channel curls around the particle and does not extend far into the channel. This contrasts with the open channel, where the flow around the particle is weaker and there is a Poiseuille flow along the whole channel length. This moving fluid column exerts a force on the second particle that gives rise to the non-decaying interaction. Figures~\ref{friction_coefficients_for_experiments_and_simulations}c and~\ref{friction_coefficients_for_experiments_and_simulations}d show a quantitative comparison between the numerical simulation and experimental data, where the solid lines indicate the simulation results and the points indicate the experimental values. The friction with the channel walls is the same for both particles, leading to the overlapping curves ($\zeta_{1,1}$ and $\zeta_{2,2}$). Meanwhile, the interaction force ($\zeta_{1,2}$) asymptotes to a non-zero value for the open channel only, similar to the correlation coefficient that was reported above. At small separations the discrepancy between simulation and experiment is likely caused by electrostatics and finite frame rate. Meanwhile, at large separations the values compare very well, with the largest discrepancy below $20\%$.
This is very good agreement given the approximations made about channel shape and width.


The agreement between the numerical simulation and the experimental data suggests that the proposed hypothesis captures the physics of non-decaying interactions between the Brownian particles. Crucially, this interaction requires flow through the ends of the channel, that is attained by keeping the two reservoirs at equal pressures. In our experiments, the pressure equalises via liquid recirculation through the secondary large channels placed far away from the narrow channel. In biological cells, this can happen through the membrane or through dedicated water transport protein channels known as aquaporins~\cite{alberts2008molecular}. In porous materials, the pressure can equalise through other interlinking channels~\cite{Karger2014}.

Our observations have far-reaching implications for diffusion processes inside channels. One of the predictions from our model is that the viscous drag on particles is smaller than the theory expects for narrow channels (figure~S5). We therefore performed an additional experiment that measured the diffusion coefficient of a single particle inside an $L=10\,\mu$m open channel (method~\cite{Dettmer2014a}). The measured diffusion coefficient is indeed $40\%$ higher than expected~\cite{Happel1973,Deen1987}: ${D_{x} / D_0 = 0.126 \pm .006}$, where $D_0$ is the diffusion coefficient in the bulk (figure~S6). This potentially explains the experimental discrepancies noted in previous studies for channels with similar dimensions~\cite{Pagliara2014a,Dettmer2014a,Schiel2014}. Because the diffusion coefficient is important for predicting transport rates, our results directly impact channel transport models~\cite{Berezhkovskii2005}. Moreover, the long-ranged interaction between particles should lead to a cooperative behaviour that could enhance transport across channels.
Also, since the interactions are not limited to two particles, they should persist for channels filled with three or even more particles. This makes our results relevant for polymers in confinement because the monomers can interact non-locally~\cite{de1979scaling}, and also for single-file systems because the particles interact not only with the closest neighbours~\cite{Wei2000}.


In conclusion, we utilised a highly controlled microfluidic system coupled with holographic optical tweezers to investigate the interaction of particles in confinement. Our measurements prove that interactions extend over the full channel length and have a constant strength that does not decay with particle separation. We explain the coupling mechanism using hydrodynamics with both a qualitative analytical model and quantitative comparisons with the numerical simulations. The excellent agreement between the theory and experiments suggests that we fully understand the properties of hydrodynamic particle interactions in microfluidic channels. The non-decaying interaction extending throughout the whole channel has important implications for the modelling of transport through channels as well as for the interpretation of experiments investigating particles diffusing in close confinement.


\begin{acknowledgments}
    The authors thank S. Ghosal and M. Muthukumar for fruitful discussions. U.F.K. was supported by an ERC starting grant (PassMembrane 261101). S.P. acknowledges funding from a Leverhulme Early Career Fellowship. K.M. was supported by a grant from the EPSRC. E.L. was supported by Marie Curie CIG grant from EU.
\end{acknowledgments}



\begin{thebibliography}{39}%
\makeatletter
\providecommand \@ifxundefined [1]{%
 \@ifx{#1\undefined}
}%
\providecommand \@ifnum [1]{%
 \ifnum #1\expandafter \@firstoftwo
 \else \expandafter \@secondoftwo
 \fi
}%
\providecommand \@ifx [1]{%
 \ifx #1\expandafter \@firstoftwo
 \else \expandafter \@secondoftwo
 \fi
}%
\providecommand \natexlab [1]{#1}%
\providecommand \enquote  [1]{``#1''}%
\providecommand \bibnamefont  [1]{#1}%
\providecommand \bibfnamefont [1]{#1}%
\providecommand \citenamefont [1]{#1}%
\providecommand \href@noop [0]{\@secondoftwo}%
\providecommand \href [0]{\begingroup \@sanitize@url \@href}%
\providecommand \@href[1]{\@@startlink{#1}\@@href}%
\providecommand \@@href[1]{\endgroup#1\@@endlink}%
\providecommand \@sanitize@url [0]{\catcode `\\12\catcode `\$12\catcode
  `\&12\catcode `\#12\catcode `\^12\catcode `\_12\catcode `\%12\relax}%
\providecommand \@@startlink[1]{}%
\providecommand \@@endlink[0]{}%
\providecommand \url  [0]{\begingroup\@sanitize@url \@url }%
\providecommand \@url [1]{\endgroup\@href {#1}{\urlprefix }}%
\providecommand \urlprefix  [0]{URL }%
\providecommand \Eprint [0]{\href }%
\providecommand \doibase [0]{http://dx.doi.org/}%
\providecommand \selectlanguage [0]{\@gobble}%
\providecommand \bibinfo  [0]{\@secondoftwo}%
\providecommand \bibfield  [0]{\@secondoftwo}%
\providecommand \translation [1]{[#1]}%
\providecommand \BibitemOpen [0]{}%
\providecommand \bibitemStop [0]{}%
\providecommand \bibitemNoStop [0]{.\EOS\space}%
\providecommand \EOS [0]{\spacefactor3000\relax}%
\providecommand \BibitemShut  [1]{\csname bibitem#1\endcsname}%
\let\auto@bib@innerbib\@empty
\bibitem [{\citenamefont {Alberts}(2008)}]{alberts2008molecular}%
  \BibitemOpen
  \bibfield  {author} {\bibinfo {author} {\bibfnamefont {B.}~\bibnamefont
  {Alberts}},\ }\href {http://books.google.co.uk/books?id=iepqmRfP3ZoC} {\emph
  {\bibinfo {title} {{Molecular Biology of the Cell: Reference Edition}}}},\
  \bibinfo {series} {Molecular Biology of the Cell: Reference Edition}\ No.\
  \bibinfo {number} {v. 1}\ (\bibinfo  {publisher} {Garland Science},\ \bibinfo
  {year} {2008})\BibitemShut {NoStop}%
\bibitem [{\citenamefont {K\"{a}rger}\ \emph {et~al.}(2014)\citenamefont
  {K\"{a}rger}, \citenamefont {Binder}, \citenamefont {Chmelik}, \citenamefont
  {Hibbe}, \citenamefont {Krautscheid}, \citenamefont {Krishna},\ and\
  \citenamefont {Weitkamp}}]{Karger2014}%
  \BibitemOpen
  \bibfield  {author} {\bibinfo {author} {\bibfnamefont {J.}~\bibnamefont
  {K\"{a}rger}}, \bibinfo {author} {\bibfnamefont {T.}~\bibnamefont {Binder}},
  \bibinfo {author} {\bibfnamefont {C.}~\bibnamefont {Chmelik}}, \bibinfo
  {author} {\bibfnamefont {F.}~\bibnamefont {Hibbe}}, \bibinfo {author}
  {\bibfnamefont {H.}~\bibnamefont {Krautscheid}}, \bibinfo {author}
  {\bibfnamefont {R.}~\bibnamefont {Krishna}}, \ and\ \bibinfo {author}
  {\bibfnamefont {J.}~\bibnamefont {Weitkamp}},\ }\href {\doibase
  10.1038/nmat3917} {\bibfield  {journal} {\bibinfo  {journal} {Nat. Mater.}\
  }\textbf {\bibinfo {volume} {13}},\ \bibinfo {pages} {333} (\bibinfo {year}
  {2014})}\BibitemShut {NoStop}%
\bibitem [{\citenamefont {Motz}\ \emph {et~al.}(2014)\citenamefont {Motz},
  \citenamefont {Schmid}, \citenamefont {H\"{a}nggi}, \citenamefont {Reguera},\
  and\ \citenamefont {Rub\'{\i}}}]{Motz2014}%
  \BibitemOpen
  \bibfield  {author} {\bibinfo {author} {\bibfnamefont {T.}~\bibnamefont
  {Motz}}, \bibinfo {author} {\bibfnamefont {G.}~\bibnamefont {Schmid}},
  \bibinfo {author} {\bibfnamefont {P.}~\bibnamefont {H\"{a}nggi}}, \bibinfo
  {author} {\bibfnamefont {D.}~\bibnamefont {Reguera}}, \ and\ \bibinfo
  {author} {\bibfnamefont {J.~M.}\ \bibnamefont {Rub\'{\i}}},\ }\href {\doibase
  10.1063/1.4892615} {\bibfield  {journal} {\bibinfo  {journal} {J. Chem.
  Phys.}\ }\textbf {\bibinfo {volume} {141}},\ \bibinfo {pages} {074104}
  (\bibinfo {year} {2014})}\BibitemShut {NoStop}%
\bibitem [{\citenamefont {Pag\`{e}s}\ \emph {et~al.}(2008)\citenamefont
  {Pag\`{e}s}, \citenamefont {James},\ and\ \citenamefont
  {Winterhalter}}]{Pages2008}%
  \BibitemOpen
  \bibfield  {author} {\bibinfo {author} {\bibfnamefont {J.-M.}\ \bibnamefont
  {Pag\`{e}s}}, \bibinfo {author} {\bibfnamefont {C.~E.}\ \bibnamefont
  {James}}, \ and\ \bibinfo {author} {\bibfnamefont {M.}~\bibnamefont
  {Winterhalter}},\ }\href {\doibase 10.1038/nrmicro1994} {\bibfield  {journal}
  {\bibinfo  {journal} {Nat. Rev. Microbiol.}\ }\textbf {\bibinfo {volume}
  {6}},\ \bibinfo {pages} {893} (\bibinfo {year} {2008})}\BibitemShut {NoStop}%
\bibitem [{\citenamefont {Deen}(1987)}]{Deen1987}%
  \BibitemOpen
  \bibfield  {author} {\bibinfo {author} {\bibfnamefont {W.}~\bibnamefont
  {Deen}},\ }\href
  {http://onlinelibrary.wiley.com/doi/10.1002/aic.690330902/full} {\bibfield
  {journal} {\bibinfo  {journal} {AIChE J.}\ }\textbf {\bibinfo {volume}
  {33}},\ \bibinfo {pages} {1409} (\bibinfo {year} {1987})}\BibitemShut
  {NoStop}%
\bibitem [{\citenamefont {Gravelle}\ \emph {et~al.}(2013)\citenamefont
  {Gravelle}, \citenamefont {Joly}, \citenamefont {Detcheverry}, \citenamefont
  {Ybert}, \citenamefont {Cottin-Bizonne},\ and\ \citenamefont
  {Bocquet}}]{Gravelle2013}%
  \BibitemOpen
  \bibfield  {author} {\bibinfo {author} {\bibfnamefont {S.}~\bibnamefont
  {Gravelle}}, \bibinfo {author} {\bibfnamefont {L.}~\bibnamefont {Joly}},
  \bibinfo {author} {\bibfnamefont {F.}~\bibnamefont {Detcheverry}}, \bibinfo
  {author} {\bibfnamefont {C.}~\bibnamefont {Ybert}}, \bibinfo {author}
  {\bibfnamefont {C.}~\bibnamefont {Cottin-Bizonne}}, \ and\ \bibinfo {author}
  {\bibfnamefont {L.}~\bibnamefont {Bocquet}},\ }\href {\doibase
  10.1073/pnas.1306447110} {\bibfield  {journal} {\bibinfo  {journal} {Proc.
  Natl. Acad. Sci. U. S. A.}\ }\textbf {\bibinfo {volume} {110}},\ \bibinfo
  {pages} {16367} (\bibinfo {year} {2013})}\BibitemShut {NoStop}%
\bibitem [{\citenamefont {Pagliara}\ \emph
  {et~al.}(2014{\natexlab{a}})\citenamefont {Pagliara}, \citenamefont
  {Dettmer},\ and\ \citenamefont {Keyser}}]{Pagliara2014}%
  \BibitemOpen
  \bibfield  {author} {\bibinfo {author} {\bibfnamefont {S.}~\bibnamefont
  {Pagliara}}, \bibinfo {author} {\bibfnamefont {S.~L.}\ \bibnamefont
  {Dettmer}}, \ and\ \bibinfo {author} {\bibfnamefont {U.~F.}\ \bibnamefont
  {Keyser}},\ }\href {\doibase 10.1103/PhysRevLett.113.048102} {\bibfield
  {journal} {\bibinfo  {journal} {Phys. Rev. Lett.}\ }\textbf {\bibinfo
  {volume} {113}},\ \bibinfo {pages} {048102} (\bibinfo {year}
  {2014}{\natexlab{a}})}\BibitemShut {NoStop}%
\bibitem [{\citenamefont {Wei}\ \emph {et~al.}(2000)\citenamefont {Wei},
  \citenamefont {Bechinger},\ and\ \citenamefont {Leiderer}}]{Wei2000}%
  \BibitemOpen
  \bibfield  {author} {\bibinfo {author} {\bibfnamefont {Q.}~\bibnamefont
  {Wei}}, \bibinfo {author} {\bibfnamefont {C.}~\bibnamefont {Bechinger}}, \
  and\ \bibinfo {author} {\bibfnamefont {P.}~\bibnamefont {Leiderer}},\ }\href
  {\doibase 10.1126/science.287.5453.625} {\bibfield  {journal} {\bibinfo
  {journal} {Science (80-. ).}\ }\textbf {\bibinfo {volume} {287}},\ \bibinfo
  {pages} {625} (\bibinfo {year} {2000})}\BibitemShut {NoStop}%
\bibitem [{\citenamefont {Dettmer}\ \emph
  {et~al.}(2014{\natexlab{a}})\citenamefont {Dettmer}, \citenamefont
  {Pagliara}, \citenamefont {Misiunas},\ and\ \citenamefont
  {Keyser}}]{Dettmer2014a}%
  \BibitemOpen
  \bibfield  {author} {\bibinfo {author} {\bibfnamefont {S.~L.}\ \bibnamefont
  {Dettmer}}, \bibinfo {author} {\bibfnamefont {S.}~\bibnamefont {Pagliara}},
  \bibinfo {author} {\bibfnamefont {K.}~\bibnamefont {Misiunas}}, \ and\
  \bibinfo {author} {\bibfnamefont {U.~F.}\ \bibnamefont {Keyser}},\ }\href
  {\doibase 10.1103/PhysRevE.89.062305} {\bibfield  {journal} {\bibinfo
  {journal} {Phys. Rev. E}\ }\textbf {\bibinfo {volume} {89}},\ \bibinfo
  {pages} {062305} (\bibinfo {year} {2014}{\natexlab{a}})}\BibitemShut
  {NoStop}%
\bibitem [{\citenamefont {Huang}\ \emph {et~al.}(2011)\citenamefont {Huang},
  \citenamefont {Chavez}, \citenamefont {Taute}, \citenamefont {Lukic},
  \citenamefont {Jeney}, \citenamefont {Raizen},\ and\ \citenamefont
  {Florin}}]{Huang2010a}%
  \BibitemOpen
  \bibfield  {author} {\bibinfo {author} {\bibfnamefont {R.}~\bibnamefont
  {Huang}}, \bibinfo {author} {\bibfnamefont {I.}~\bibnamefont {Chavez}},
  \bibinfo {author} {\bibfnamefont {K.~M.}\ \bibnamefont {Taute}}, \bibinfo
  {author} {\bibfnamefont {B.}~\bibnamefont {Lukic}}, \bibinfo {author}
  {\bibfnamefont {S.}~\bibnamefont {Jeney}}, \bibinfo {author} {\bibfnamefont
  {M.~G.}\ \bibnamefont {Raizen}}, \ and\ \bibinfo {author} {\bibfnamefont
  {E.-L.}\ \bibnamefont {Florin}},\ }\href {\doibase 10.1038/nphys1953}
  {\bibfield  {journal} {\bibinfo  {journal} {Nat. Phys.}\ }\textbf {\bibinfo
  {volume} {7}} (\bibinfo {year} {2011}),\ 10.1038/nphys1953},\ \Eprint
  {http://arxiv.org/abs/1003.1980} {arXiv:1003.1980} \BibitemShut {NoStop}%
\bibitem [{\citenamefont {Happel}\ and\ \citenamefont
  {Brenner}(1973)}]{Happel1973}%
  \BibitemOpen
  \bibfield  {author} {\bibinfo {author} {\bibfnamefont {J.}~\bibnamefont
  {Happel}}\ and\ \bibinfo {author} {\bibfnamefont {H.}~\bibnamefont
  {Brenner}},\ }\href@noop {} {\emph {\bibinfo {title} {{Low Reynolds number
  hydrodynamics}}}},\ edited by\ \bibinfo {editor} {\bibfnamefont
  {R.}~\bibnamefont {Moreau}}\ (\bibinfo  {publisher} {Martinus Nijhoff
  Publishers},\ \bibinfo {year} {1973})\BibitemShut {NoStop}%
\bibitem [{\citenamefont {Diamant}(2009)}]{Diamant2009}%
  \BibitemOpen
  \bibfield  {author} {\bibinfo {author} {\bibfnamefont {H.}~\bibnamefont
  {Diamant}},\ }\href {\doibase 10.1143/JPSJ.78.041002} {\bibfield  {journal}
  {\bibinfo  {journal} {J. Phys. Soc. Japan}\ }\textbf {\bibinfo {volume}
  {78}},\ \bibinfo {pages} {1} (\bibinfo {year} {2009})},\ \Eprint
  {http://arxiv.org/abs/0812.4971} {arXiv:0812.4971} \BibitemShut {NoStop}%
\bibitem [{\citenamefont {Crocker}(1997)}]{Crocker1997}%
  \BibitemOpen
  \bibfield  {author} {\bibinfo {author} {\bibfnamefont {J.~C.}\ \bibnamefont
  {Crocker}},\ }\href@noop {} {\bibfield  {journal} {\bibinfo  {journal} {J.
  Chem. Phys.}\ }\textbf {\bibinfo {volume} {106}},\ \bibinfo {pages} {2837}
  (\bibinfo {year} {1997})}\BibitemShut {NoStop}%
\bibitem [{\citenamefont {Cui}\ \emph {et~al.}(2004)\citenamefont {Cui},
  \citenamefont {Diamant}, \citenamefont {Lin},\ and\ \citenamefont
  {Rice}}]{Cui2004}%
  \BibitemOpen
  \bibfield  {author} {\bibinfo {author} {\bibfnamefont {B.}~\bibnamefont
  {Cui}}, \bibinfo {author} {\bibfnamefont {H.}~\bibnamefont {Diamant}},
  \bibinfo {author} {\bibfnamefont {B.}~\bibnamefont {Lin}}, \ and\ \bibinfo
  {author} {\bibfnamefont {S.}~\bibnamefont {Rice}},\ }\href {\doibase
  10.1103/PhysRevLett.92.258301} {\bibfield  {journal} {\bibinfo  {journal}
  {Phys. Rev. Lett.}\ }\textbf {\bibinfo {volume} {92}},\ \bibinfo {pages}
  {258301} (\bibinfo {year} {2004})}\BibitemShut {NoStop}%
\bibitem [{\citenamefont {Cui}\ \emph {et~al.}(2002)\citenamefont {Cui},
  \citenamefont {Diamant},\ and\ \citenamefont {Lin}}]{Cui2002}%
  \BibitemOpen
  \bibfield  {author} {\bibinfo {author} {\bibfnamefont {B.}~\bibnamefont
  {Cui}}, \bibinfo {author} {\bibfnamefont {H.}~\bibnamefont {Diamant}}, \ and\
  \bibinfo {author} {\bibfnamefont {B.}~\bibnamefont {Lin}},\ }\href {\doibase
  10.1103/PhysRevLett.89.188302} {\bibfield  {journal} {\bibinfo  {journal}
  {Phys. Rev. Lett.}\ }\textbf {\bibinfo {volume} {89}},\ \bibinfo {pages}
  {188302} (\bibinfo {year} {2002})}\BibitemShut {NoStop}%
\bibitem [{\citenamefont {Valley}\ \emph {et~al.}(2007)\citenamefont {Valley},
  \citenamefont {Rice}, \citenamefont {Cui}, \citenamefont {Ho}, \citenamefont
  {Diamant},\ and\ \citenamefont {Lin}}]{Valley2007a}%
  \BibitemOpen
  \bibfield  {author} {\bibinfo {author} {\bibfnamefont {D.~T.}\ \bibnamefont
  {Valley}}, \bibinfo {author} {\bibfnamefont {S.~a.}\ \bibnamefont {Rice}},
  \bibinfo {author} {\bibfnamefont {B.}~\bibnamefont {Cui}}, \bibinfo {author}
  {\bibfnamefont {H.~M.}\ \bibnamefont {Ho}}, \bibinfo {author} {\bibfnamefont
  {H.}~\bibnamefont {Diamant}}, \ and\ \bibinfo {author} {\bibfnamefont
  {B.}~\bibnamefont {Lin}},\ }\href {\doibase 10.1063/1.2719191} {\bibfield
  {journal} {\bibinfo  {journal} {J. Chem. Phys.}\ }\textbf {\bibinfo {volume}
  {126}},\ \bibinfo {pages} {134908} (\bibinfo {year} {2007})}\BibitemShut
  {NoStop}%
\bibitem [{\citenamefont {Kosheleva}\ \emph {et~al.}(2012)\citenamefont
  {Kosheleva}, \citenamefont {Leahy}, \citenamefont {Diamant}, \citenamefont
  {Lin},\ and\ \citenamefont {Rice}}]{Kosheleva2012}%
  \BibitemOpen
  \bibfield  {author} {\bibinfo {author} {\bibfnamefont {E.}~\bibnamefont
  {Kosheleva}}, \bibinfo {author} {\bibfnamefont {B.}~\bibnamefont {Leahy}},
  \bibinfo {author} {\bibfnamefont {H.}~\bibnamefont {Diamant}}, \bibinfo
  {author} {\bibfnamefont {B.}~\bibnamefont {Lin}}, \ and\ \bibinfo {author}
  {\bibfnamefont {S.~a.}\ \bibnamefont {Rice}},\ }\href {\doibase
  10.1103/PhysRevE.86.041402} {\bibfield  {journal} {\bibinfo  {journal} {Phys.
  Rev. E}\ }\textbf {\bibinfo {volume} {86}},\ \bibinfo {pages} {041402}
  (\bibinfo {year} {2012})}\BibitemShut {NoStop}%
\bibitem [{\citenamefont {Liron}\ and\ \citenamefont
  {Shahar}(1978)}]{Liron1978}%
  \BibitemOpen
  \bibfield  {author} {\bibinfo {author} {\bibfnamefont {N.}~\bibnamefont
  {Liron}}\ and\ \bibinfo {author} {\bibfnamefont {R.}~\bibnamefont {Shahar}},\
  }\href@noop {} {\bibfield  {journal} {\bibinfo  {journal} {J. Fluid Mech.}\
  }\textbf {\bibinfo {volume} {86}} (\bibinfo {year} {1978})}\BibitemShut
  {NoStop}%
\bibitem [{\citenamefont {{Al Quddus}}\ \emph {et~al.}(2008)\citenamefont {{Al
  Quddus}}, \citenamefont {Moussa},\ and\ \citenamefont
  {Bhattacharjee}}]{AlQuddus2008}%
  \BibitemOpen
  \bibfield  {author} {\bibinfo {author} {\bibfnamefont {N.}~\bibnamefont {{Al
  Quddus}}}, \bibinfo {author} {\bibfnamefont {W.~a.}\ \bibnamefont {Moussa}},
  \ and\ \bibinfo {author} {\bibfnamefont {S.}~\bibnamefont {Bhattacharjee}},\
  }\href {\doibase 10.1016/j.jcis.2007.09.060} {\bibfield  {journal} {\bibinfo
  {journal} {J. Colloid Interface Sci.}\ }\textbf {\bibinfo {volume} {317}},\
  \bibinfo {pages} {620} (\bibinfo {year} {2008})}\BibitemShut {NoStop}%
\bibitem [{\citenamefont {Frydel}\ and\ \citenamefont
  {Diamant}(2010)}]{Frydel2010}%
  \BibitemOpen
  \bibfield  {author} {\bibinfo {author} {\bibfnamefont {D.}~\bibnamefont
  {Frydel}}\ and\ \bibinfo {author} {\bibfnamefont {H.}~\bibnamefont
  {Diamant}},\ }\href {\doibase 10.1103/PhysRevLett.104.248302} {\bibfield
  {journal} {\bibinfo  {journal} {Phys. Rev. Lett.}\ }\textbf {\bibinfo
  {volume} {104}},\ \bibinfo {pages} {248302} (\bibinfo {year}
  {2010})}\BibitemShut {NoStop}%
\bibitem [{\citenamefont {Reguera}\ \emph {et~al.}(2006)\citenamefont
  {Reguera}, \citenamefont {Schmid}, \citenamefont {Burada}, \citenamefont
  {Rub\'{\i}}, \citenamefont {Reimann},\ and\ \citenamefont
  {H\"{a}nggi}}]{Reguera2006}%
  \BibitemOpen
  \bibfield  {author} {\bibinfo {author} {\bibfnamefont {D.}~\bibnamefont
  {Reguera}}, \bibinfo {author} {\bibfnamefont {G.}~\bibnamefont {Schmid}},
  \bibinfo {author} {\bibfnamefont {P.}~\bibnamefont {Burada}}, \bibinfo
  {author} {\bibfnamefont {J.}~\bibnamefont {Rub\'{\i}}}, \bibinfo {author}
  {\bibfnamefont {P.}~\bibnamefont {Reimann}}, \ and\ \bibinfo {author}
  {\bibfnamefont {P.}~\bibnamefont {H\"{a}nggi}},\ }\href {\doibase
  10.1103/PhysRevLett.96.130603} {\bibfield  {journal} {\bibinfo  {journal}
  {Phys. Rev. Lett.}\ }\textbf {\bibinfo {volume} {96}},\ \bibinfo {pages}
  {130603} (\bibinfo {year} {2006})}\BibitemShut {NoStop}%
\bibitem [{\citenamefont {Pagliara}\ \emph {et~al.}(2011)\citenamefont
  {Pagliara}, \citenamefont {Chimerel}, \citenamefont {Langford}, \citenamefont
  {Aarts},\ and\ \citenamefont {Keyser}}]{Pagliara2011}%
  \BibitemOpen
  \bibfield  {author} {\bibinfo {author} {\bibfnamefont {S.}~\bibnamefont
  {Pagliara}}, \bibinfo {author} {\bibfnamefont {C.}~\bibnamefont {Chimerel}},
  \bibinfo {author} {\bibfnamefont {R.}~\bibnamefont {Langford}}, \bibinfo
  {author} {\bibfnamefont {D.~G. a.~L.}\ \bibnamefont {Aarts}}, \ and\ \bibinfo
  {author} {\bibfnamefont {U.~F.}\ \bibnamefont {Keyser}},\ }\href {\doibase
  10.1039/c1lc20399a} {\bibfield  {journal} {\bibinfo  {journal} {Lab Chip}\
  }\textbf {\bibinfo {volume} {11}},\ \bibinfo {pages} {3365} (\bibinfo {year}
  {2011})}\BibitemShut {NoStop}%
\bibitem [{\citenamefont {Grier}(2003)}]{Grier2003}%
  \BibitemOpen
  \bibfield  {author} {\bibinfo {author} {\bibfnamefont {D.~G.}\ \bibnamefont
  {Grier}},\ }\href {\doibase 10.1038/nature01935} {\bibfield  {journal}
  {\bibinfo  {journal} {Nature}\ }\textbf {\bibinfo {volume} {424}},\ \bibinfo
  {pages} {810} (\bibinfo {year} {2003})}\BibitemShut {NoStop}%
\bibitem [{\citenamefont {Padgett}\ and\ \citenamefont {{Di
  Leonardo}}(2011)}]{Padgett2011}%
  \BibitemOpen
  \bibfield  {author} {\bibinfo {author} {\bibfnamefont {M.}~\bibnamefont
  {Padgett}}\ and\ \bibinfo {author} {\bibfnamefont {R.}~\bibnamefont {{Di
  Leonardo}}},\ }\href {\doibase 10.1039/c0lc00526f} {\bibfield  {journal}
  {\bibinfo  {journal} {Lab Chip}\ }\textbf {\bibinfo {volume} {11}},\ \bibinfo
  {pages} {1196} (\bibinfo {year} {2011})}\BibitemShut {NoStop}%
\bibitem [{\citenamefont {Pagliara}\ \emph {et~al.}(2013)\citenamefont
  {Pagliara}, \citenamefont {Schwall},\ and\ \citenamefont
  {Keyser}}]{Pagliara2013}%
  \BibitemOpen
  \bibfield  {author} {\bibinfo {author} {\bibfnamefont {S.}~\bibnamefont
  {Pagliara}}, \bibinfo {author} {\bibfnamefont {C.}~\bibnamefont {Schwall}}, \
  and\ \bibinfo {author} {\bibfnamefont {U.~F.}\ \bibnamefont {Keyser}},\
  }\href {\doibase 10.1002/adma.201203500} {\bibfield  {journal} {\bibinfo
  {journal} {Adv. Mater.}\ }\textbf {\bibinfo {volume} {25}},\ \bibinfo {pages}
  {844} (\bibinfo {year} {2013})}\BibitemShut {NoStop}%
\bibitem [{\citenamefont {Dettmer}\ \emph
  {et~al.}(2014{\natexlab{b}})\citenamefont {Dettmer}, \citenamefont {Keyser},\
  and\ \citenamefont {Pagliara}}]{Dettmer2014}%
  \BibitemOpen
  \bibfield  {author} {\bibinfo {author} {\bibfnamefont {S.~L.}\ \bibnamefont
  {Dettmer}}, \bibinfo {author} {\bibfnamefont {U.~F.}\ \bibnamefont {Keyser}},
  \ and\ \bibinfo {author} {\bibfnamefont {S.}~\bibnamefont {Pagliara}},\
  }\href {\doibase 10.1063/1.4865552} {\bibfield  {journal} {\bibinfo
  {journal} {Rev. Sci. Instrum.}\ }\textbf {\bibinfo {volume} {85}},\ \bibinfo
  {pages} {023708} (\bibinfo {year} {2014}{\natexlab{b}})}\BibitemShut
  {NoStop}%
\bibitem [{\citenamefont {Riley}\ \emph {et~al.}(2006)\citenamefont {Riley},
  \citenamefont {Hobson},\ and\ \citenamefont {Bence}}]{riley2006mathematical}%
  \BibitemOpen
  \bibfield  {author} {\bibinfo {author} {\bibfnamefont {K.~F.}\ \bibnamefont
  {Riley}}, \bibinfo {author} {\bibfnamefont {P.}~\bibnamefont {Hobson}}, \
  and\ \bibinfo {author} {\bibfnamefont {S.~J.}\ \bibnamefont {Bence}},\ }\href
  {http://books.google.co.uk/books?id=Mq1nlEKhNcsC} {\emph {\bibinfo {title}
  {{Mathematical Methods for Physics and Engineering: A Comprehensive
  Guide}}}}\ (\bibinfo  {publisher} {Cambridge University Press},\ \bibinfo
  {year} {2006})\BibitemShut {NoStop}%
\bibitem [{\citenamefont {Dufresne}\ \emph {et~al.}(2000)\citenamefont
  {Dufresne}, \citenamefont {Squires}, \citenamefont {Brenner},\ and\
  \citenamefont {Grier}}]{Dufresne2000}%
  \BibitemOpen
  \bibfield  {author} {\bibinfo {author} {\bibfnamefont {E.~R.}\ \bibnamefont
  {Dufresne}}, \bibinfo {author} {\bibfnamefont {T.~M.}\ \bibnamefont
  {Squires}}, \bibinfo {author} {\bibfnamefont {M.~P.}\ \bibnamefont
  {Brenner}}, \ and\ \bibinfo {author} {\bibfnamefont {D.~G.}\ \bibnamefont
  {Grier}},\ }\href {http://www.ncbi.nlm.nih.gov/pubmed/11019330} {\bibfield
  {journal} {\bibinfo  {journal} {Phys. Rev. Lett.}\ }\textbf {\bibinfo
  {volume} {85}},\ \bibinfo {pages} {3317} (\bibinfo {year}
  {2000})}\BibitemShut {NoStop}%
\bibitem [{\citenamefont {Mazo}(2002)}]{Mazo2002}%
  \BibitemOpen
  \bibfield  {author} {\bibinfo {author} {\bibfnamefont {R.~M.}\ \bibnamefont
  {Mazo}},\ }\href@noop {} {\emph {\bibinfo {title} {{Brownian Motion:
  Fluctuations, Dynamics, and Applications}}}}\ (\bibinfo  {publisher}
  {Clarendon press, Oxford University Press},\ \bibinfo {address} {Oxford},\
  \bibinfo {year} {2002})\BibitemShut {NoStop}%
\bibitem [{\citenamefont {Smythe}(1961)}]{Smythe1961}%
  \BibitemOpen
  \bibfield  {author} {\bibinfo {author} {\bibfnamefont {W.~R.}\ \bibnamefont
  {Smythe}},\ }\href {\doibase 10.1063/1.1706394} {\bibfield  {journal}
  {\bibinfo  {journal} {Phys. Fluids}\ }\textbf {\bibinfo {volume} {4}},\
  \bibinfo {pages} {756} (\bibinfo {year} {1961})}\BibitemShut {NoStop}%
\bibitem [{\citenamefont {Batchelor}(2000)}]{batchelor2000introduction}%
  \BibitemOpen
  \bibfield  {author} {\bibinfo {author} {\bibfnamefont {G.~K.}\ \bibnamefont
  {Batchelor}},\ }\href {http://books.google.co.uk/books?id=Rla7OihRvUgC}
  {\emph {\bibinfo {title} {{An Introduction to Fluid Dynamics}}}},\ Cambridge
  Mathematical Library\ (\bibinfo  {publisher} {Cambridge University Press},\
  \bibinfo {year} {2000})\BibitemShut {NoStop}%
\bibitem [{\citenamefont {Xu}\ \emph {et~al.}(2005)\citenamefont {Xu},
  \citenamefont {Rice}, \citenamefont {Lin},\ and\ \citenamefont
  {Diamant}}]{Xu2005}%
  \BibitemOpen
  \bibfield  {author} {\bibinfo {author} {\bibfnamefont {X.}~\bibnamefont
  {Xu}}, \bibinfo {author} {\bibfnamefont {S.}~\bibnamefont {Rice}}, \bibinfo
  {author} {\bibfnamefont {B.}~\bibnamefont {Lin}}, \ and\ \bibinfo {author}
  {\bibfnamefont {H.}~\bibnamefont {Diamant}},\ }\href {\doibase
  10.1103/PhysRevLett.95.158301} {\bibfield  {journal} {\bibinfo  {journal}
  {Phys. Rev. Lett.}\ }\textbf {\bibinfo {volume} {95}},\ \bibinfo {pages}
  {158301} (\bibinfo {year} {2005})}\BibitemShut {NoStop}%
\bibitem [{\citenamefont {Pagliara}\ \emph
  {et~al.}(2014{\natexlab{b}})\citenamefont {Pagliara}, \citenamefont
  {Dettmer}, \citenamefont {Misiunas}, \citenamefont {Lea}, \citenamefont
  {Tan},\ and\ \citenamefont {Keyser}}]{Pagliara2014a}%
  \BibitemOpen
  \bibfield  {author} {\bibinfo {author} {\bibfnamefont {S.}~\bibnamefont
  {Pagliara}}, \bibinfo {author} {\bibfnamefont {S.~L.}\ \bibnamefont
  {Dettmer}}, \bibinfo {author} {\bibfnamefont {K.}~\bibnamefont {Misiunas}},
  \bibinfo {author} {\bibfnamefont {L.}~\bibnamefont {Lea}}, \bibinfo {author}
  {\bibfnamefont {Y.}~\bibnamefont {Tan}}, \ and\ \bibinfo {author}
  {\bibfnamefont {U.~F.}\ \bibnamefont {Keyser}},\ }\href {\doibase
  10.1140/epjst/e2014-02324-6} {\bibfield  {journal} {\bibinfo  {journal} {Eur.
  Phys. J. Spec. Top.}\ }\textbf {\bibinfo {volume} {223}},\ \bibinfo {pages}
  {3145} (\bibinfo {year} {2014}{\natexlab{b}})}\BibitemShut {NoStop}%
\bibitem [{\citenamefont {Schiel}\ and\ \citenamefont
  {Siwy}(2014)}]{Schiel2014}%
  \BibitemOpen
  \bibfield  {author} {\bibinfo {author} {\bibfnamefont {M.}~\bibnamefont
  {Schiel}}\ and\ \bibinfo {author} {\bibfnamefont {Z.~S.}\ \bibnamefont
  {Siwy}},\ }\href {\doibase 10.1021/jp505823r} {\bibfield  {journal} {\bibinfo
   {journal} {J. Phys. Chem. C}\ }\textbf {\bibinfo {volume} {118}},\ \bibinfo
  {pages} {19214} (\bibinfo {year} {2014})}\BibitemShut {NoStop}%
\bibitem [{\citenamefont {Berezhkovskii}\ and\ \citenamefont
  {Bezrukov}(2005)}]{Berezhkovskii2005}%
  \BibitemOpen
  \bibfield  {author} {\bibinfo {author} {\bibfnamefont {A.~M.}\ \bibnamefont
  {Berezhkovskii}}\ and\ \bibinfo {author} {\bibfnamefont {S.~M.}\ \bibnamefont
  {Bezrukov}},\ }\href {\doibase 10.1529/biophysj.104.057588} {\bibfield
  {journal} {\bibinfo  {journal} {Biophys. J.}\ }\textbf {\bibinfo {volume}
  {88}},\ \bibinfo {pages} {L17} (\bibinfo {year} {2005})}\BibitemShut
  {NoStop}%
\bibitem [{\citenamefont {{De Gennes}}(1979)}]{de1979scaling}%
  \BibitemOpen
  \bibfield  {author} {\bibinfo {author} {\bibfnamefont {P.-G.}\ \bibnamefont
  {{De Gennes}}},\ }\href@noop {} {\emph {\bibinfo {title} {{Scaling concepts
  in polymer physics}}}}\ (\bibinfo  {publisher} {Cornell university press},\
  \bibinfo {year} {1979})\BibitemShut {NoStop}%
\bibitem [{\citenamefont {Brenner}(1990)}]{Supplementary}%
  \BibitemOpen
  See Supplementary Material [url], which includes Refs. [38-40]
  \BibitemShut {NoStop}%
\bibitem [{\citenamefont {Brenner}(1990)}]{Brenner1990}%
  \BibitemOpen
  \bibfield  {author} {\bibinfo {author} {\bibfnamefont {H.}~\bibnamefont
  {Brenner}},\ }\href@noop {} {\bibfield  {journal} {\bibinfo  {journal}
  {Langmuir}\ }\textbf {\bibinfo {volume} {6}} (\bibinfo {year}
  {1990})}\BibitemShut {NoStop}%
\bibitem [{\citenamefont {Ermak}\ and\ \citenamefont
  {McCammon}(1978)}]{Ermak1978}%
  \BibitemOpen
  \bibfield  {author} {\bibinfo {author} {\bibfnamefont {D.~L.}\ \bibnamefont
  {Ermak}}\ and\ \bibinfo {author} {\bibfnamefont {J.~a.}\ \bibnamefont
  {McCammon}},\ }\href {\doibase 10.1063/1.436761} {\bibfield  {journal}
  {\bibinfo  {journal} {J. Chem. Phys.}\ }\textbf {\bibinfo {volume} {69}},\
  \bibinfo {pages} {1352} (\bibinfo {year} {1978})}\BibitemShut {NoStop}%
\bibitem [{\citenamefont {Savin}\ and\ \citenamefont
  {Doyle}(2005)}]{Savin2005}%
  \BibitemOpen
  \bibfield  {author} {\bibinfo {author} {\bibfnamefont {T.}~\bibnamefont
  {Savin}}\ and\ \bibinfo {author} {\bibfnamefont {P.~S.}\ \bibnamefont
  {Doyle}},\ }\href {\doibase 10.1529/biophysj.104.042457} {\bibfield
  {journal} {\bibinfo  {journal} {Biophys. J.}\ }\textbf {\bibinfo {volume}
  {88}},\ \bibinfo {pages} {623} (\bibinfo {year} {2005})}\BibitemShut
  {NoStop}%
\end{thebibliography}

%

\end{document}